\documentclass[aps,
pre,
superscriptaddress,
amsmath,amssymb,
reprint,
showkeys,
]{revtex4-2}

\usepackage{algorithm}
\usepackage{algpseudocode}
\usepackage{array}
\usepackage{mathtools}
\usepackage[dvipsnames]{xcolor}
\usepackage{lipsum}
\usepackage{tikz}
\usepackage{wrapfig}
\usepackage[T1]{fontenc}
\usepackage[utf8]{inputenc}
\usepackage{amsmath}
\usepackage{amssymb}
\usepackage{graphicx}
\usepackage{refstyle}
\usepackage{multirow}
\usepackage{blkarray,booktabs,bigstrut}
\usepackage{placeins}

\usepackage[
citecolor=blue,
colorlinks,
linkcolor=blue,
urlcolor=blue,
]{hyperref}
\usepackage{makecell}
\usepackage{xcolor}
\usepackage{dcolumn}
\usepackage{bm}
\usepackage{tabularx,booktabs}
\usepackage{multirow}
\usepackage{float}
\usepackage[normalem]{ulem}
\usepackage{multirow}
\usepackage{blkarray}

\begin{document}

	\title{ Stochastic ultimatum game: Spite-driven resource feedback fosters fairness}
	
	\author{Arunava Patra}
	\email{arunava20@iitk.ac.in}
	\address{
		Department of Physics,
		Indian Institute of Technology Kanpur,
		Uttar Pradesh 208016, India
	}
	\author{Prosanta Mandal}
	\email{prosantam21@iitk.ac.in (corresponding author)}
	\address{
		Department of Physics,
		Indian Institute of Technology Kanpur,
		Uttar Pradesh 208016, India
	}
	
	\author{Sagar Chakraborty}
	\email{sagarc@iitk.ac.in}
	\address{
		Department of Physics,
		Indian Institute of Technology Kanpur,
		Uttar Pradesh 208016, India
	}

\begin{abstract}
Resource scarcity can fundamentally encourage antisocial behaviour, whereas resource abundance can promote fair behaviour. Experimental evidence indeed suggests that scarcity induces spiteful behaviour, while repeated interactions enhance fairness. However, existing studies of game--environment feedback systems are largely confined to the evolution of cooperation and they overlook the interplay between resources, spite, and fairness. To address this lacuna, we develop a stochastic ultimatum game framework in which an offerer and an accepter repeatedly interact to negotiate exploitation of a self-renewable resource under the ownership of the offerer. Successful agreements deplete the resource, whereas unsuccessful agreements inhibit exploitation and facilitate replenishment. The mutation--selection driven two-species stochastic evolutionary dynamics reveal that the emergence of spite and fairness strongly depends on the resource growth rate. Fairness predominantly prevails for resources with high growth rates. Intriguingly, low resource growth rates give rise to a resource feedback loop driven by spite: spiteful behaviour dominates in the depleted state, facilitating transition of the resource state to replete state which, in turn, promotes fairness through repeated interactions.
\end{abstract}

\maketitle
\section{Introduction}
  Organisms inevitably compete for resources in order to reap greater benefits and enhance their chances of survival. Resource scarcity can intensify competition, resulting in the emergence of antisocial behaviours such as spite~\cite{PVH_2014_JPE}---an act that harms other organisms while incurring a personal cost~\cite{H_1970_NAT, KP_1979_NAT, FRW_2000_TEE, GW_2004_JEB, GW_2004_Sc, HBL_2010_EVO}. Since spiteful behaviour entails a cost to individuals but provides no direct benefit, its existence in biological systems appears to be evolutionarily disadvantageous. Nevertheless, natural selection favours spiteful behaviour~\cite{D_1860_BOOK}, as it has been observed across taxa ranging from microbes to humans. Among non-human species, evidence of spite has been reported in certain bacteria~\cite{HBL_2010_EVO}, parasitoid wasps~\cite{GHT_2007_AME_NAT}, and red fire ants~\cite{KR_1998_NAT}. This gives rise to a fundamental evolutionary puzzle: How does natural selection promote the evolution of spite?
  
 Understanding the evolutionary origin of spite is particularly important because antisocial acts can, paradoxically, facilitate the emergence and maintenance of social behaviour within a population. In particular, spiteful behaviour may contribute to sustaining fairness~\cite{FS_2014_PRSB}. Fairness---a social behaviour involving the equitable distribution of resources---contradicts the rational outcome at the individual level and is also not found to be evolutionarily stable~\cite{RTON_2013_PNAS, DBA_2016_EBV, FS_2014_PRSB}. Nevertheless, a growing body of experimental and empirical evidences indicate that the tendency toward fair behaviour extends beyond humans and is present in several non-human species, including chimpanzees, brown capuchin monkeys, and domestic dogs~\cite{H_1999_BOOK, BW_2003_NAT, B_2004_BP, BSW_2005_PRSB, B_2006_SJR, RHVH_2009_PNAS, M_2010_ZT}. Moreover, a sense of fairness has also been empirically documented in coyotes and ravens~\cite{B_2004_BP, H_1999_BOOK}. Uncovering the mechanisms underlying the emergence of fairness has also attracted increasing interdisciplinary attention because of its potential implications for social justice~\cite{B_2005_NJ}, equitable treatment~\cite{BW_2014_Sc}, and fair resource distribution among individuals~\cite{DBA_2017_PO}. Altogether, a comprehensive investigation into the evolutionary connection between spite and fairness in an evolving resource environment is therefore warranted.
 
 A promising framework for investigating such evolutionary phenomena is evolutionary game theory~\cite{SP_1973_NAT, M_1974_JTB, S_1988_BOOK}. Within this framework, the fitness associated with different behaviors is modeled using games, while their evolution is governed by the principles of natural selection. In particular, both spite and fairness are frequently studied using the ultimatum game~\cite{NPS_2000_Sc, YLW_2015_EL, HMB_2006_Sc, HCS_2017_PNAS}. 
 The ultimatum game provides a standard framework for resource division between two individuals. In this game, a \emph{desirable} (e.g, food or money) is given to an individual, termed offerer, who proposes a division with another individual, termed accepter. The accepter can either accept or reject the offer. Accepting the offer leads to a successful \emph{agreement}, allowing both individuals to obtain their respective shares, whereas rejection prevents the agreement and leaves both individuals empty-handed. An equal division of the desirable is typically interpreted as fairness, while an unequal division is regarded as unfairness. Offering an extremely low share while demanding a high share is often considered a manifestation of spiteful behaviour, since such actions substantially reduce the chance of a successful agreement, thereby imposing a cost on both individuals~\cite{FS_2014_PRSB, PM_1996_OBHDP}.  
 
 A variety of mechanisms have been proposed over the years to explain the emergence of fair behaviour in the context of the UG, including reputation updating through repeated interactions~\cite{S_1999_Experimental}, network reciprocity~\cite{PNS_2000_PRSB, A_2007_BOOK, KS_2001_PRSB}, variation in stake size~\cite{WCW_2014_SR, ZCL_2018_AMC}, learning dynamics~\cite{GBS_1995_GEB}, rare mutation processes, and finite population effects~\cite{RTON_2013_PNAS}. Furthermore, psychological attributes such as empathy and emotion have also been identified as potential drivers of fairness~\cite{P_2002_BMB, TS_2016_PO}. In contrast, the evolutionary origins of spite are relatively less understood. Existing studies suggest that factors such as dynamic networks~\cite{FFS_2021_NATCOM}, finite population size~\cite{SF_2012_EVOL}, indirect reciprocity~\cite{JB_2004_PRSB}, negative assortment~\cite{FS_2014_PRSB}, negative relatedness~\cite{PVH_2014_JPE}, and prejudice~\cite{PZC_2025_PRE} can facilitate the evolution of spiteful behaviour. In addition, emotions such as envy and anger have been shown to promote spiteful behaviour in experiments~\cite{Wobker2014, Pillutla1996}.
 
 The aforementioned studies overlook several aspects of realistic settings. In particular, limited availability of natural resources constitutes a threat to individuals~\cite{H_1968_Sc, O_2008_PMUK}, while their uneven distribution can intensify conflicts among them. Furthermore, ownership is recognized as a widespread phenomenon in biological systems and is defined as a situation in which an owner possesses a resource and others respect that possession~\cite{HRB_2016_JEB}. Interestingly, ownership plays a crucial role in preventing the tragedy of the commons by reducing unregulated competition among individuals~\cite{CO_2010_EJ}. Instead of attempting to seize resources directly, nonowners may obtain benefits through negotiation with the owner, for example by contributing labour, expertise, or harvesting ability in exchange for a share of the extracted output.
 
 Owners and nonowners may engage in repeated interactions as part of a negotiation process. Within the framework of the UG, owners and nonowners can be identified as offerers and accepters, respectively. Such repeated interactions may proceed in various ways. In particular, repeated encounters between the same offerer and accepter over multiple rounds provide the accepter with an opportunity to punish unfair behaviour spitefully in expectation of receiving more favourable offers in future interactions. It is also shown experimentally that repeated interactions with the same opponent can promote fair behaviour in the UG~\cite{S_1999_Experimental}.

 Integrating ownership, repeated interactions, and evolving resources into a unified framework may provide theoretical insights into how ownership and resource growth shape the coevolution of spite and fairness, and how this coevolution, in turn, influences resource dynamics. Moreover, this formulation allows us to develop a game--environment feedback framework in the context of the UG. Notably, prior frameworks for modelling repeated interactions and game--environment feedback, namely the stochastic game framework~\cite{HSC_2018_NAT, S_1953_PNAS}, have mostly focused on the Prisoner's Dilemma~\cite{RC_1965}. Therefore, adapting the stochastic game framework to the UG is particularly important, as it may facilitate future investigations into how learning and memory shape the evolution of spite and fairness in a fluctuating environment.
 
We therefore introduce the UG in a stochastic game setting in which the offerer owns a resource and repeatedly interacts with the accepter to negotiate resource exploitation. In every round, the desirable---over which the UG is being played---is a fixed fraction of resource.  In that round, depending on her action, the offerer offers a fraction of the desirable to the accepter. Subsequently, completing the round, the accepter may either accept or reject the offer depending on her action. In the event of acceptance, the entire desirable is deemed harvested from the resource, thereby depleting the resource. On the other hand, the rejection by the accepter prevents exploitation the desirable out of the resource and hence resource level stays unchanged. Additionally, as we have considered in this paper, if the resource is self-renewing one, its level can change independent of the exploitation by the players.  The changing resource level, in turn, also modulates the desirable, the offers, and the demands; in effect, the underlying payoffs of the UG game may change round-by-round giving rise to a stochastic game. We aim to examine the consequences of resource level dynamics on the evolution of spite and fairness, and investigate the evolutionary dynamics under a \emph{mutation--selection regime}~\cite{FI_2006_JET, IN_2009_ProcRSB}.
 
\section{Two-player Stochastic mini-UG}

We consider an \emph{alternating} repeated interaction~\cite{PNH_2022_NATCOM}  between an offerer and an accepter aiming to divide a shared resource, where the offerer owns the resource and the accepter demands a share in exchange for her expertise in extracting it. In an alternating interaction, one player moves first and the other player moves afterward in each round. Generically, the player who moves first is called the leader, while the other player is called the follower. The leader, adopting a general memory-$\frac{\kappa}{2}$ strategy~\cite{CLT_2017_IEEECS, PD_2012_PNAS, HS_1997_PRSB, HMVC_2017_PNAS}, decides the action in a given round based on the previous $\kappa$ number of actions, including her own and the follower's. A memory-$\frac{\kappa}{2}$ follower also decides the action based on the previous $\kappa$ actions of both herself and the leader. Since the leader moves first and the follower acts afterward in a given round, the sets of previous $\kappa$ actions are different for the leader and the follower. While the set of previous $\kappa$ actions---on which the leader bases her action---consists of the actions exclusively from the previous rounds, the set of previous $\kappa$ actions---on which the follower bases her action---includes the leader's action in the current round.

Naturally, the offerer acts as the leader, while the accepter responds to the offerer's actions in the alternating interaction. The strategic interaction underlying their repeated encounters is best represented by the UG. We assume that the offerer and the accepter adopt the cognitively least demanding memory-$\frac{1}{2}$ strategy, also known as a \emph{reactive strategy}~\cite{NS_1990_AAM, HS_1998_BOOK}, because our goal is to construct the simplest possible non-trivial framework; considering memory-$\frac{\kappa}{2}$ strategies with higher $\kappa$ would substantially increase the complexity of the model. Moreover, using the UG as the underlying game can further complicate the formulation of reactive strategies, since defining such strategies would require infinitely many parameters. Therefore, for analytical tractability, we take the mini-UG as the underlying game.

In the mini-UG~\cite{PZC_2025_PRE}, the offerer may choose to make either a high or a low offer, while the accepter may choose to demand either a high or a low share in order to divide a desirable normalized to unity. The high offer and high demand correspond to an equal split of the resource and are therefore viewed as a fair offer and a fair demand, respectively. By contrast, the low offer and low demand correspond to a minimal share and are considered an unfair offer and an unfair demand, respectively. We denote the actions \emph{high} and \emph{low} by H and L, respectively, while the corresponding offer or demand amounts are denoted by $h$ and $l$. These parameters satisfy $0 < l < h=0.5$. 

When the offerer selects action H, an agreement is reached regardless of the accepter's action. Consequently, both action pairs (H, H) and (H, L) result in successful agreements, yielding payoffs of $1-h$ to the offerer and $h$ to the accepter. The action pair (H, H) represents \emph{fairness} as both individuals favor an equal division of the resource. Meanwhile, the action pair (H, L) represents the \emph{altruistic interaction}. In contrast, when the offerer chooses action L, an agreement is reached only if the accepter also chooses L. The resulting action pair (L, L) corresponds to \emph{unfairness}, where the offerer receives the larger share, $1-l$, and the accepter receives only $l$. By contrast, the action pair (L, H) corresponds to a \emph{spiteful interaction}, in which no agreement is reached and both individuals receive zero payoff. This outcome is interpreted as spiteful because a low offer and a high demand cause both individuals to forgo their own payoffs in an attempt to reduce the other's gain.

  \begin{figure}[h]
	\centering
	\includegraphics[scale=0.48]{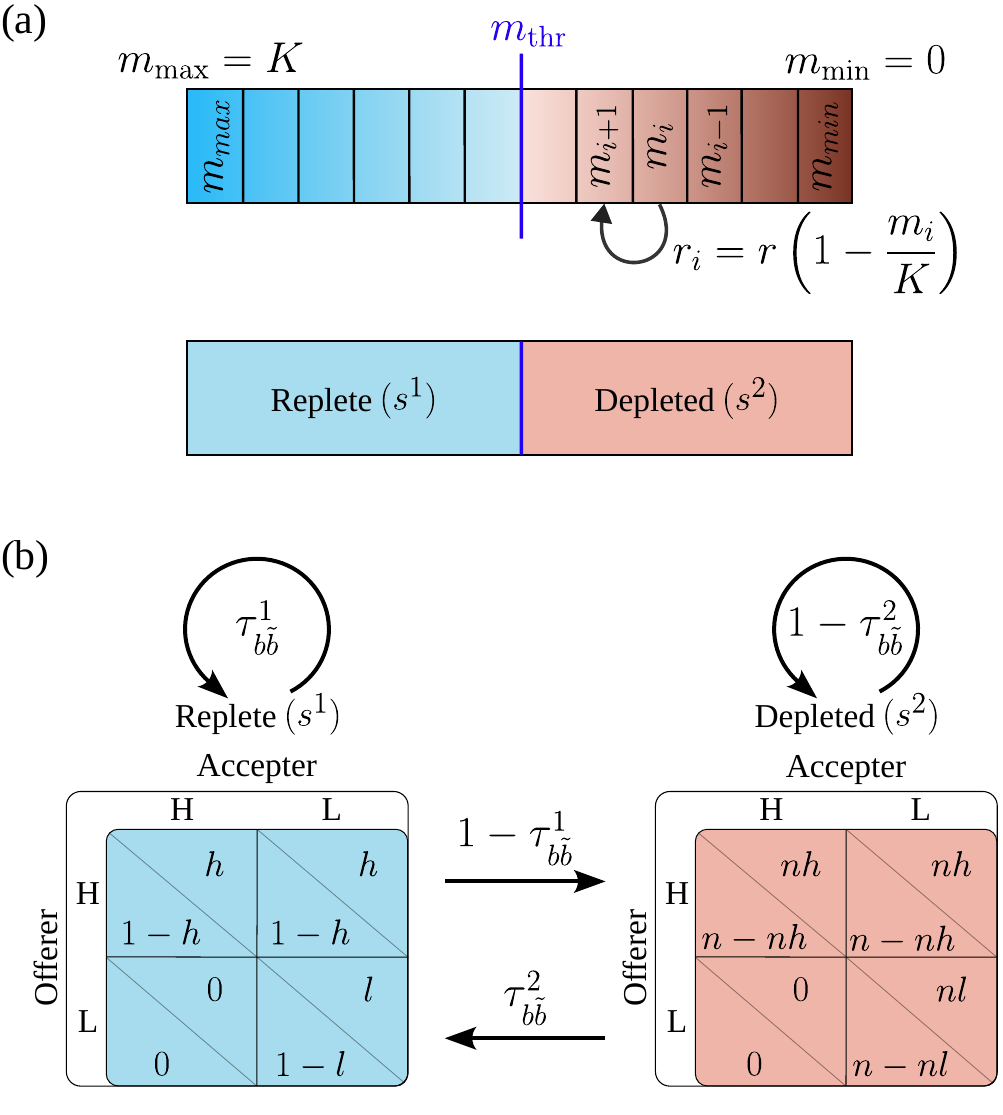}
	\caption{Schematic diagram illustrating the coarse-graining of a continuous resource into a two-state resource system and the transitions between the states. Subplot~(a) demonstrates the discretization of the continuous resource into two states: the replete state, denoted by $(s^1)$, and the depleted state, denoted by $(s^2)$. The discrete resource levels are represented by $m_i$, with $m_{\text{max}}=K$, where $K$ is the carrying capacity of the resource. The growth rate at resource level $m_i$ is denoted by $r_i$ and is governed by the logistic equation. Resource levels above the threshold $m_{\text{thr}}$ are classified as replete, whereas those below it are classified as depleted. Subplot~(b) presents the underlying mini-UG in the replete (light blue payoff matrix) and depleted (light brown payoff matrix) states, along with the transitions between them. The component $\tau^{i}_{b\tilde{b}}$ of the transition vector $\bm{\tau}$ represents the transition probability from state $s^i$ to $s^1$ for the action pair $(b,\tilde{b})$ of the offerer and the accepter. Here, $n$ represents the relative abundance of the depleted state with respect to the replete state.}
	\label{fig:two_state_reduction}
  \end{figure}

The desirable---the amount extracted from the resource following a successful agreement---depends on the resource abundance. The abundance of the resource is most naturally represented by a continuous variable, and the dynamics of a \emph{self-renewing resource} can be represented by the logistic growth equation~\cite{B_2011_SL, M_2002_IAM}, characterized by an intrinsic growth rate $r>0$ and a carrying capacity $K>0$. However, incorporating a continuous resource into a stochastic game framework substantially increases analytical complexity. To obtain a simple and tractable model, we therefore discretize the resource into $M+1$ levels, denoted by $m_i$ with $i\in{0,1,\ldots,M}$. Here, $m_0=m_{\min}=0$ represents the minimum resource abundance, while $m_M=m_{\max}=K$ corresponds to the carrying capacity. The transition rate from level $m_i$ to $m_{i+1}$ is determined by the logistic equation, $ r_i=r\left(1-\frac{m_i}{K}\right)$.
 We further reduce the model to a two-state resource system by introducing an abundance threshold $m_{\mathrm{thr}}$. Resource levels satisfying $m_i>m_{\mathrm{thr}}$ are grouped into a single \emph{replete} state, $s^1$, while all remaining levels constitute a \emph{depleted} state, $s^2$. The reduction of the continuous resource to a two-state resource system is depicted in Fig.~\ref{fig:two_state_reduction}(a).
   
Since we consider a two-state resource system, the extracted amounts from the resource after a successful agreement in the replete and depleted states are $n_1$ and $n_2$, respectively. Naturally, these quantities satisfy $(n_1 > n_2)$ due to the fact that individuals should exploit more resources in the replete state than in the depleted state. As the total amount of resource (per round) divided between the offerer and the accepter is normalized to unity in mini-UG, we set the amount of resource exploited  (per round) in the replete state to unity, i.e., $(n_1 = 1)$. The corresponding amount in the depleted state is then denoted by $(n \equiv n_2)$, where $(0 < n < 1)$. Consequently, strategic interactions in the replete and depleted states are governed by two distinct mini-UGs, as illustrated in Fig.~\ref{fig:two_state_reduction}(b).

\subsection{Reactive strategy in stochastic mini-UG}
A reactive strategy in a two-state stochastic game consists of four conditional probabilities since the reactive strategy becomes resource state dependent. We represent the offerer's reactive strategy by ${\bm S}_o=(p_o^1,q_o^1;p_o^2,q_o^2),$ where $p_o^i$ and $q_o^i$ denote the probabilities of choosing action H in state $s^i$ given that the accepter's previous action was H and L, respectively.  Similarly, the accepter's reactive strategy is represented by ${\bm S}_a=(p_a^1,q_a^1;p_a^2,q_a^2),$ where $p_a^i$ and $q_a^i$ denote the probabilities of choosing action H in state $s^i$ given that the offerer's previous action was H and L, respectively. Note that in the very first round, as there is no existing opponent-action to respond to, the offerer must be assigned an unconditional probability of choosing H in the initial round depending on the resource state.

In our setup, in line with previous work~\cite{MPC_2026_arX} on the repeated simultaneous mini-UG, we consider only \emph{pure reactive strategies}, characterized by $p_v^i,q_v^i\in\{0,1\}$, where $v\in\{o,a\}$. The pure reactive strategy $(0,0)$ in a given state corresponds to an unfair strategy in that state, denoted by $\text{U}_o$ for the offerer and $\text{U}_a$ for the accepter. By contrast, in any resource state, the reactive strategy $(1,1)$ is a \emph{fair strategy}, denoted by $\text{F}_o$ and $\text{F}_a$, respectively. The strategy $(1,0)$ is a \emph{complier strategy}: the offerer with this strategy complies with the accepter's previous demand, while the accepter with this strategy complies with the offerer's previous offer. We denote these strategies by $\text{C}_o$ and $\text{C}_a$. Conversely, the strategy $(0,1)$ is an \emph{anti-complier strategy}: the offerer anti-complies with the accepter's previous demand, while the accepter anti-complies with the offerer's previous offer. These strategies are denoted by $\text{A}_o$ and $\text{A}_a$. Irrespective of the resource state, the offerer's initial action is taken to be H for all pure reactive strategies except for the unfair strategy where the initial action is taken to be L. It is obvious that notationally the pure strategies can conveniently be denoted using the symbols---F, C, U and A: e.g., the pure reactive strategy $(1,1;1,0)$ can be represented by $(\text{F}_o;\text{C}_o)$ for the offerer and $(\text{F}_a;\text{C}_a)$ for the accepter. For ready reference, we tabulate all the pure reactive strategies in Fig.~\ref{fig:Transition_with_strategies}(a).

\subsection{Transition vectors in stochastic mini-UG}

While transitions between resource states influence the strategy choices of the offerer and the accepter, their actions also affect the stochastic transitions between resource states~\cite{HSC_2018_NAT, KHS_2023_NATCOM}. The effect of an action pair on resource dynamics is described by the transition vector $\bm{\tau}=\left(\tau^1_{\mathrm{HH}},\tau^1_{\mathrm{HL}},\tau^1_{\mathrm{LH}},\tau^1_{\mathrm{LL}};\tau^2_{\mathrm{HH}},\tau^2_{\mathrm{HL}},\tau^2_{\mathrm{LH}},\tau^2_{\mathrm{LL}}\right)$, where $\tau^i_{b\tilde{b}}$ denotes the probability that the next resource state is the replete state $s^1$, given that the current state is $s^i$, the offerer chooses $b\in\{\mathrm{H},\mathrm{L}\}$, and the accepter chooses $\tilde{b}\in\{\mathrm{H},\mathrm{L}\}$. When all elements satisfy $\tau^i_{b\tilde{b}}\in\{0,1\}$, the transition vector is deterministic. We restrict our attention to deterministic transition vectors, as they are sufficient to capture the essential features of the resource dynamics.

There are $2^8$ possible deterministic transition vectors. However, as we shall argue below that owing to the facts that successful agreements allow the offerer and the accepter to exploit the resource (thereby reducing its abundance) and unsuccessful agreements prevent exploitation  (thereby allowing the self-renewing resource to recover), the number of non-trivial transition vectors relevant for our work in this paper can be reduced to only two which correspond to two different growth rates. 

First we observe that since agreements are successful for the action pairs (H, H), (H, L), and (L, L), the transition probabilities corresponding to these action pairs in a given state must be identical. Therefore, the transition vector can be written in the form $\bm{\tau}=\left(\tau^1,\tau^1,\tau'^1,\tau^1;\tau^2,\tau^2,\tau'^2,\tau^2\right)$, reducing the total number of possible transition vectors to $2^4$. Furthermore, transition vectors of the form $\bm{\tau}=\left(\tau^1,\tau^1,0,\tau^1;\tau^2,\tau^2,\tau'^2,\tau^2\right)$ are not admissible because they represent (note here $\tau^{1}_{\text{LH}}=0$) the unrealistic fact that unsuccessful agreements [due to action pair (L, H)] in replete state leads to resource-exploitation. Also, transition vectors of the form $\bm{\tau}=\left(\tau^1,\tau^1,\tau'^1,\tau^1;1,1,0,1\right)$ are again irrelevant because, if successful agreements replenish the resource in depleted state, an unsuccessful agreement must also replenish it; in symbols, if $\tau^2_{\mathrm{HH}}=\tau^2_{\mathrm{HL}}=\tau^2_{\mathrm{LL}}=1$, then $\tau^2_{\mathrm{LH}}\ne 0$.

Now, within the aforementioned two constraints, we consider only those self-renewing resources for which the action pairs may lead to non-trivial dynamics where continual transitions between the two resource states may be effected. Accordingly, we first eliminate the transition vector $\bm{\tau}=\left(1,1,1,1;1,1,1,1\right)$ from consideration as here the system will be only in the replete state. We also eliminate the transition vector $\bm{\tau}=\left(0,0,1,0;0,0,0,0\right)$ because the resource never returns to the replete state once it becomes depleted. For the similar reasons, the transition vectors $\bm{\tau}=\left(1,1,1,1;0,0,1,0\right)$ and $\bm{\tau}=\left(1,1,1,1;0,0,0,0\right)$ are also excluded. In conclusion, the only remaining relevant transition vectors that should be of our interest are $\bm{\tau}_{1111}\equiv\left(0,0,1,0;1,1,1,1\right)$ and $\bm{\tau}_{0010}\equiv\left(0,0,1,0;0,0,1,0\right)$.
 
The interpretation of ${\bm \tau}_{0010}$ and ${\bm \tau}_{1111}$ in terms of self renewing resource is more subtle. It may seem puzzling that, under ${\bm \tau}_{1111}$, any action pair in the depleted state replenishes the resource, whereas action pairs corresponding to successful agreements in the replete state degrade it. This apparently counter-intuitive feature can be understood by associating ${\bm \tau}_{1111}$ with a resource having a suitable high $r$ (say, $r_{1111}$). The growth rate ($r_i$, a decreasing function of $m_i$) becomes vanishingly small when the resource abundance is close to its carrying capacity $(m_i\sim K)$. Consequently, successful agreements can have a substantial impact on a replete resource. In contrast, when the resource is moderately depleted ($0\ll m_i<m_{\mathrm{thr}}$), its growth rate is sufficiently large to restore it to the replete state even without any unsuccessful agreement. Thus, the resource can recover despite continued exploitation.

A similar interpretation applies to the transition vector ${\bm \tau}_{0010}$. In this case, successful agreements in the replete state also have a similar impact and drive the resource to the depleted state. However, in contrast to the transition vector ${\bm \tau}_{1111}$, successful agreements in the depleted state also have a substantial impact and keep the resource depleted. It occurs if the resource growth rate is not sufficient to outcompete exploitation when the resource is intermediately depleted ($0\ll m_i<m_{\mathrm{thr}}$). This is expected in the scenario, if  $r_{1111}\gg r_{0010}$, where $r_{0010}$ denotes the intrinsic growth rate of the resource with ${\bm \tau}_{0010}$. Thus, ${\bm \tau}_{0010}$ represents a comparatively slowly-growing resource, whereas ${\bm \tau}_{1111}$ corresponds to a comparatively faster-growing resource. 

\begin{figure*}[ht]
	\centering
	\includegraphics[scale=0.48]{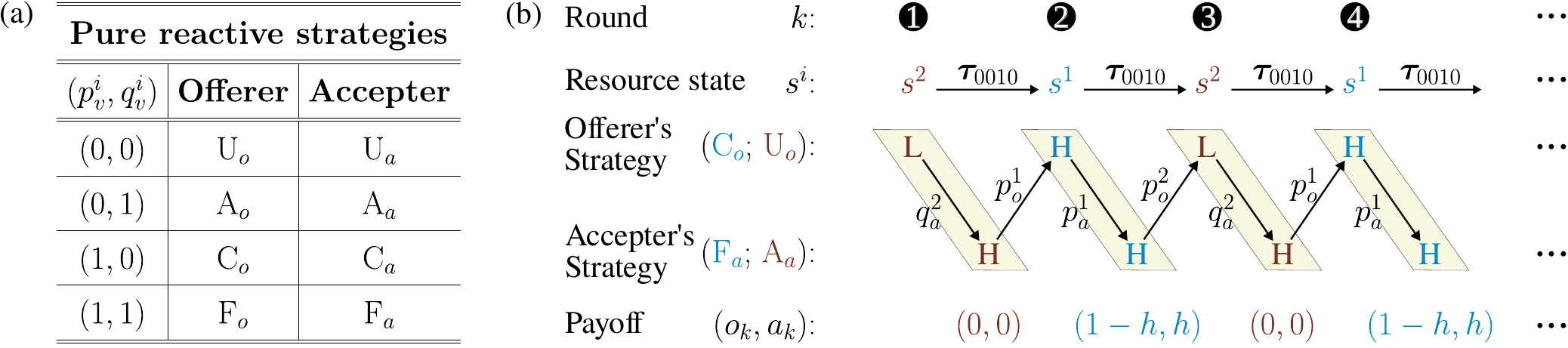}
	\caption{Schematic representation of the two-state stochastic game: The pure reactive strategies in state $s^i$ are tabulated in subplot (a), where $(p_v^i,q_v^i)$ represents the reactive strategy in state $s^i$. The subscript $v$ can be either $o$ or $a$, depending on whether the player is the offerer or the accepter. Subplot (b): In the schematic presentation of the alternating game with resource-feedback, the strategies of the offerer and the accepter are taken to be $(\text{C}_o;\text{U}_o)$ and $(\text{F}_a;\text{A}_a)$, respectively, for illustrative purpose. In addition, the transition vector governing the resource transition is taken to be ${\bm \tau}_{0010}\equiv\left(\tau^1_{\mathrm{HH}}=0,\tau^1_{\mathrm{HL}}=0,\tau^1_{\mathrm{LH}}=1,\tau^1_{\mathrm{LL}}=0;\tau^2_{\mathrm{HH}}=0,\tau^2_{\mathrm{HL}}=0,\tau^2_{\mathrm{LH}}=1,\tau^2_{\mathrm{LL}}=0\right)$. Since the repeated interaction begins in the depleted state, the offerer moves first and plays the action L as her strategy is $\text{U}_o$ in the depleted state. The accepter then plays the action H with probability $q_a^2=1$, as her strategy in the depleted state is $\text{A}_a$. Consequently, both the offerer and the accepter receive zero payoffs. In the next round, the resource transitions to the replete state $s^1$ because $\tau^2_{\mathrm{LH}}=1$ [note the action pair is (L, H) in depleted state ($s^2$) in the first round]. Subsequently, the offerer complies with the accepter's previous action with probability $p_o^1=1$. The accepter then follows the offerer's current action H with probability $p_a^1=1$. In that round, the offerer receives the payoff $1-h$, while the accepter receives $h$. In the third round, the resource transitions back to the depleted state because $\tau^1_{\mathrm{HH}}=0$ [note the action pair in the second round is (H, H) in replete state $s^2$],  and the repeated interaction proceeds in a similar manner ad infinitum.}
		\label{fig:Transition_with_strategies}
\end{figure*}

To crystal clearly illustrate the \emph{resource-game feedback}, we provide a schematic representation of the stochastic game between the offerer and the accepter in Fig.~\ref{fig:Transition_with_strategies}.

\subsection{Average payoffs}
At the end of a repeated play, the offerer and the accepter obtain accumulated payoffs consisting of the payoffs obtained at each round during their repeated interactions. Although the interaction may continue indefinitely, the likelihood that they interact again depends on the \emph{shadow of the future}~\cite{BRT_2015_SR, B_2005_AER, CF_2015_JECP}: a longer shadow corresponds to a higher probability of future interaction. To account for this, payoffs are discounted by a factor $\delta\in(0,1)$, known as the \emph{discount factor}~\cite{GV_2016_JET}. Equivalently, $\delta$ can be interpreted as the probability that the interaction continues to the next round. Hence, the probability of occurring $k$-th round is $\delta^{k-1}$.

The average payoff is obtained by normalizing the accumulated discounted payoff received over the course of repeated interactions by the expected number of rounds, $1/(1-\delta)$. Let ${o}_k$ and ${a}_k$ denote the payoffs of the offerer and the accepter, respectively, in the $k$-th round. The average payoffs of an offerer using strategy $\bm{S}_o$ and an accepter using strategy $\bm{S}_a$ are given by
\begin{subequations}
	\begin{eqnarray}
		\pi_o({\bm{S}_o}, \bm{S}_a)=(1-\delta)\sum_{k=1}^{\infty} {o}_{k}\delta^{k-1},
		\label{eq:Average_cum_payoff_offerer}\\
		\pi_a({\bm{S}_o}, \bm{S}_a)=(1-\delta)\sum_{k=1}^{\infty} {a}_{k}\delta^{k-1},
		\label{eq:Average_cum_payoff_accepter}
	\end{eqnarray}
\end{subequations}
where ${o}_k\in \{1-h, 0, 1-l, n(1-h), n(1-l)\}$ and  ${a}_k\in \{h, 0, l, nh, nl\}$.

\section{Two-species stochastic evolutionary dynamics}
In order to examine the long-term (mutation-selection regime) stochastic evolution of fairness and spite, we consider a two-species population of the offerers and the accepters with subpopulation sizes $N_o$ and $N_a$, respectively. We assume that each subpopulation is monomorphic. Since mutation is an inevitable part of the evolutionary process, we consider the rare mutation process. Under this mutation process, a mutant either fixates or becomes extinct before the next mutation occurs. The population state changes only through mutant fixation.

In this setting, a population state is characterized by the strategies adopted by the offerer and the accepter subpopulations. Let $\bm{S}_o$ and $\bm{S}_a$ denote the strategies of the offerer and the accepter subpopulations, respectively. Then, the pair $(\bm{S}_o,\bm{S}_a)$ uniquely specifies the \emph{population state}. In a given population state, mutations may arise either in one subpopulation or simultaneously in both subpopulations. When mutants appear in both subpopulations, three outcomes are possible after the fixation process. The mutants may either fail to fixate in either resident subpopulation, successfully fixate in both of them, or fixate in only one of the two subpopulations. For example, a mutant pair $(\tilde{\bm{S}}_o,\tilde{\bm{S}}_a)$ in a population with state $(\bm{S}_o,\bm{S}_a)$ produces three new population states $(\tilde{\bm{S}}_o,\tilde{\bm{S}}_a)$, $(\bm{S}_o,\tilde{\bm{S}}_a)$, and $(\tilde{\bm{S}}_o,\bm{S}_a)$. Consequently, the resident population may transition to any one of three new population states or remain in the current state.

It should be noted that unlike existing asymmetric game models~\cite{VH_2016_JET} that limit mutations to a single subpopulation at a time, we permit simultaneous mutations in both subpopulations as done in previous work~\cite{MPC_2026_arX} on simultaneous repeated mini-UG, where this adjustment ensured consistency with the definition of a two-species evolutionarily stable strategy, which demands robustness against any mutant invasion, single or concurrent. To reflect the fact that joint mutations are much less common, our model automatically and consistently weights their occurrences as the product of the mutation rates of both subpopulations (see Appendix~\ref{appen:Analytical framework of two-species long-term stochastic evolutionary dynamics}).


At each time step, a mutant pair arises in a population consisting of two monomorphic subpopulations and either fixates or goes extinct, thereby restoring the population to a state with two monomorphic subpopulations. Subsequently, another mutant pair may appear. In this manner, the mutation--selection process continues over an infinite time, {leading to successive transitions between population states} over a long evolutionary timescale. Mutants in the offerer and the accepter subpopulations are drawn from their respective strategy sets at rates $\mu_o$ and $\mu_a$. Since we restrict attention to pure reactive strategies, the strategy set of the offerer is given by $$\mathbb{S}_o \equiv \{\bm{S}_o=(S_o^1; S_o^2)\,|\, S_o^1, S_o^2 \in \{\text{U}_o, \text{F}_o, \text{A}_o, \text{C}_o\}\},$$ and that of the accepter by $$\mathbb{S}_a \equiv \{ \bm{S}_a=(S_a^1; S_a^2)\,|\, S_a^1, S_a^2 \in \{\text{U}_a, \text{F}_a, \text{A}_a, \text{C}_a\}\}.$$

The period from the appearance of a mutant to its fixation or extinction constitutes the transient phase. The competition between the mutant and the resident in the transient phase is governed by the replication--selection process, which ultimately determines the mutant's fate in terms of its fixation probability---the probability that the mutant takes over the entire population. This probability typically depends on the strength of selection, denoted by $w$, which measures the contribution of payoff to fitness. Under weak selection, the contribution of payoff to fitness is small, and random drift dominates the evolutionary dynamics. Details of the calculation of fixation probabilities are provided in Appendix~\ref{appen:Birth--death process in two-species population}.

Let $\nu(\bm{S}_o, \bm{S}_a)$ and $\gamma(\bm{S}_o, \bm{S}_a)$ denote the spiteful and fairness rates, respectively, of a population in state $(\bm{S}_o, \bm{S}_a)$. The quantities $\nu(\bm{S}_o, \bm{S}_a)$ and $\gamma(\bm{S}_o, \bm{S}_a)$ measure the frequencies with which the action pairs $(\text{L}, \text{H})$ and $(\text{H}, \text{H})$, respectively, arise in repeated interactions between an offerer using strategy $\bm{S}_o$ and an accepter using strategy $\bm{S}_a$. Both quantities evolve over time as the population state changes. Denoting by $\alpha^{(t)}{(\bm{S}_o,\bm{S}_a)}$ the abundance of state $(\bm{S}_o,\bm{S}_a)$ at time $t$, the average spiteful and fairness rates can be written as
\begin{eqnarray}
	\langle \nu(t) \rangle = \sum_{\substack{ S_o\in\mathcal{S}_o\\S_a\in\mathcal{S}_a}} \alpha_{({S}_{o}, {S}_{a})}^{(t)}\, \nu(\bm{S}_o, \bm{S}_a),
	\label{eq:avg_spiteful_rate}\\
	\langle \gamma(t) \rangle = \sum_{\substack{ S_o\in\mathcal{S}_o\\S_a\in\mathcal{S}_a}} \alpha_{({S}_{o}, {S}_{a})}^{(t)}\, \gamma(\bm{S}_o, \bm{S}_a),
	\label{eq:avg_fairness_rate}
\end{eqnarray}
It is also important to find out the probability of resource being in the replete state. We denote $\beta(\bm{S}_o,\bm{S}_a)$ as the probability of the resource being in the replete state when the population state is $(\bm{S}_o,\bm{S}_a)$. Then, the average replete resource level can be expressed as follows:
\begin{eqnarray}
	\langle \beta(t) \rangle = \sum_{\substack{ S_o\in\mathcal{S}_o\\S_a\in\mathcal{S}_a}} \alpha_{({S}_{o}, {S}_{a})}^{(t)}\, \beta(\bm{S}_o, \bm{S}_a),
	\label{eq:replete_resource_level}
\end{eqnarray}
Detailed calculation of $\nu(\bm{S}_o, \bm{S}_a)$ and $\gamma(\bm{S}_o, \bm{S}_a)$ are provided in Appendix~\ref{appen:Fairness and spite rate of reactive strategy in two-player stochastic mini-UG}, while the computation of $\alpha^{(t)}_{(\bm{S}_o, \bm{S}_a)}$ is described in Appendix~\ref{appen:Analytical framework of two-species long-term stochastic evolutionary dynamics}.

\section{Results}

\begin{figure}[h]
	\centering
	\includegraphics[scale=0.35]{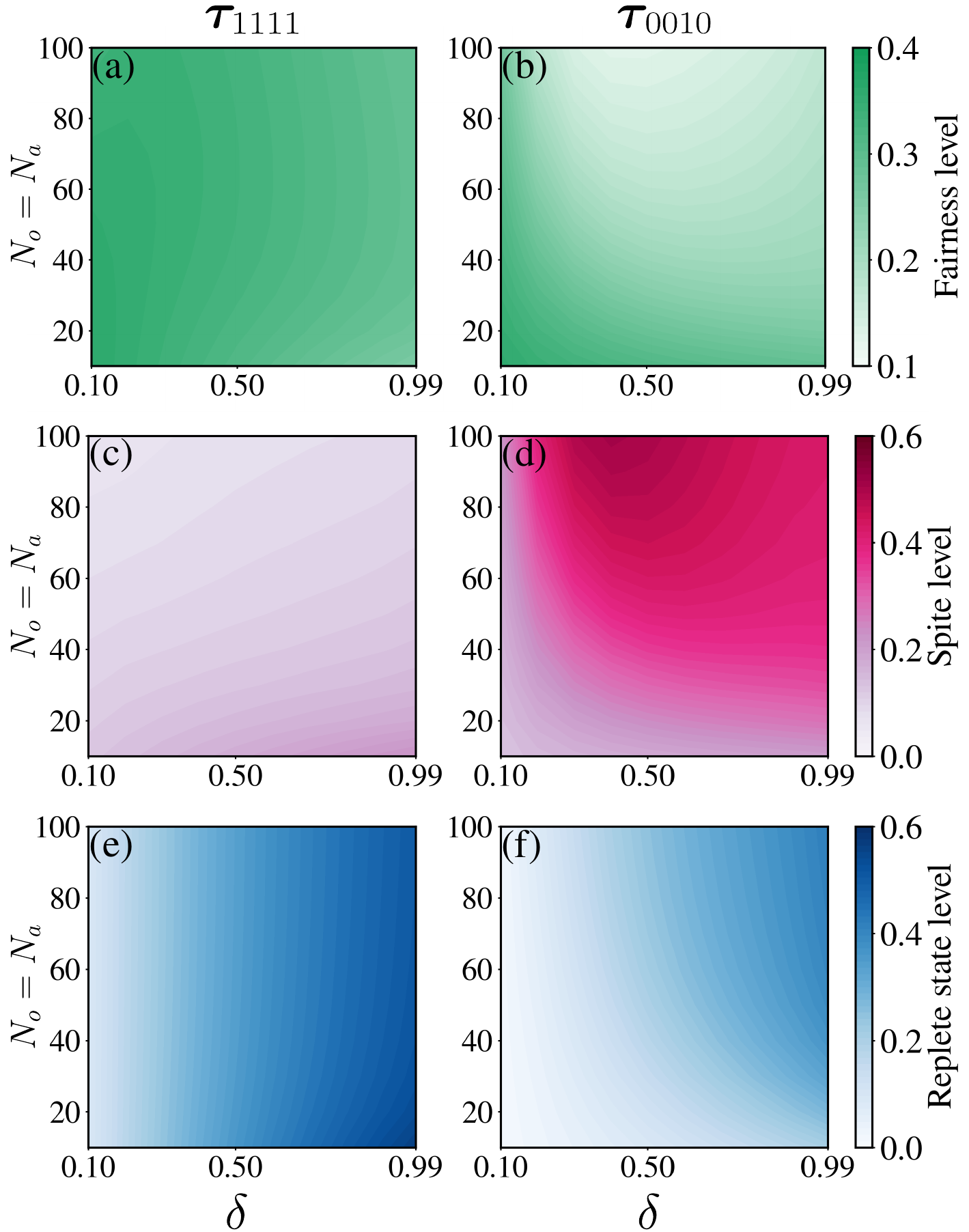}
	\caption{The fairness, spite, and replete state levels are shown across different discount factors and population sizes for two transition vectors. The first and second columns correspond to the transition vectors ${\bm \tau}_{0100}$ and ${\bm \tau}_{1111}$, respectively. The discount factor and population sizes are chosen from the intervals $\delta \in [0.1,0.99]$ and $N_o=N_a\in[10,100]$, respectively. Subplots (a) and (b) in the first row show the fairness levels, and subplots (c) and (d) in the second row illustrate the spite levels. The replete levels are shown in subplots (e) and (f) in the third row. The color gradient from light to dark indicates increasing values of the corresponding level.}
	\label{fig:spite_fair_resource_level}
\end{figure}

To investigate the evolution of spite and fairness in a mutation--selection process, we consider equal-sized offerer and accepter subpopulations. Furthermore, the mutation rates of the two subpopulations are assumed to be equal and fixed at $\mu_o=\mu_a=10^{-2}$, while the intensity of selection is set to $w=0.5$. The high offer is chosen as $h=0.5$, since an equal division of the resource represents fair behavior. In contrast, the low offer is fixed at the minimum possible share, namely $l=0.05$. We then examine the long-term levels of spite, fairness, and replete resources across discount factors $\delta\in[0.1,0.99]$ and population sizes $N_o=N_a\in[10,100]$ for the two transition vectors ${\bm \tau}_{0010}$ and ${\bm \tau}_{1111}$, as shown in Fig.~\ref{fig:spite_fair_resource_level}.

To generate Fig.~\ref{fig:spite_fair_resource_level}, we evolve the system for $10^5$ generations from an initial population state consisting of unfair offerer and accepter subpopulations. We then record the distribution of population states after $10^5$ generations and compute the average spite level using Eq.~\ref{eq:avg_spiteful_rate}, the average fairness level using Eq.~\ref{eq:avg_fairness_rate}, and the replete resource level using Eq.~\ref{eq:replete_resource_level} for each combination of discount factor and population size. Note that the initial resource state in the repeated interaction is taken to be the depleted state.

Fig.~\ref{fig:spite_fair_resource_level} shows that the levels of fairness and spite are strongly influenced by the transition vector. For the transition vectors ${\bm \tau}_{1111}$, the fairness level remains high and varies little across population sizes and discount factors. By contrast, for the transition vector ${\bm \tau}_{0010}$, the fairness level exhibits substantial variation across both parameters [see Figs.~\ref{fig:spite_fair_resource_level}(a)--\ref{fig:spite_fair_resource_level}(b)]. A similar pattern is observed for spite, although the spite level remains comparatively lower for ${\bm \tau}_{1111}$ [see Figs.~\ref{fig:spite_fair_resource_level}(c)--\ref{fig:spite_fair_resource_level}(d)].  For high discount factor and small population,  the replete resource level is low for the transition vector ${\bm \tau}_{0010}$ whereas it is high for the transition vector ${\bm \tau}_{1111}$.

Under ${\bm \tau}_{1111}$, the resource always transitions from the depleted state to the replete state, irrespective of the action pair. As a result, the offerer and the accepter mostly interact in the replete state, leading to a high replete state level. Moreover, the replete state level increases with the discount factor because longer game length provide more opportunities for the resource to recover from its initially depleted state; see Fig.~\ref{fig:spite_fair_resource_level}(e). Since interactions under ${\bm \tau}_{1111}$ occur predominantly in the replete state, the repeated game effectively reduce to repeated play within a single underlying mini-UG. Just as repeated simultaneous interactions in a single mini-UG promote fairness~\cite{MPC_2026_arX}, it is not surprising then that fairness emerges for this transition vector.

For the transition vector ${\bm \tau}_{0010}$, the levels of spite, fairness, and replete state remain largely unchanged with increasing discount factor for small populations. In contrast, these quantities vary substantially with the discount factor for large populations. In particular, the spite level increases with the discount factor, attains its maximum at intermediate discount factors, and then decreases slightly. The fairness level exhibits the opposite behavior; compare Figs.~\ref{fig:spite_fair_resource_level}(b) and \ref{fig:spite_fair_resource_level}(d). At the same time, the replete resource level increases monotonically with the discount factor. The contrasting behaviors observed in small and large populations can be attributed to the relative roles of random drift and selection. In small populations, random drift dominates the evolutionary dynamics and largely obscures the effects of game payoffs. As population size increases, however, selection becomes more effective, allowing payoff differences generated by repeated interactions to shape the evolutionary outcome. The increase in the replete resource level with the discount factor arises because the resource is initially depleted and longer interactions increase the level of spite, thereby facilitating resource replenishment. 

To understand the relationship between spite, fairness, and resource abundance for the transition vector ${\bm \tau}_{0010}$, we examine the asymptotic distribution of population states in the mutation--selection process. For this purpose, we classify the $2^8$ population states---each characterized by  strategy a pair---according to the outcomes they generate in the replete and depleted states. For convenience, we abbreviate fair, unfair, altruistic, and spiteful outcomes as FO, UO, AO, and SO, respectively, and use subscripts---`rep' and `dep'---to indicate the corresponding resource state is replete or
depleted. A fair outcome in a given resource state is defined to have occurred when a strategy pair exclusively generates the action pair (H, H) in that state. Similarly, an unfair outcome is defined to have occurred when only the action pair (L, L) is generated repeatedly. Along the same line, a spiteful outcome corresponds to the exclusive occurrence of (L, H), whereas an altruistic outcome corresponds to the exclusive occurrence of (H, L). The above definitions of outcomes do not take into account the action pair played in the first round.

\begin{figure}[h]
	\centering
	\includegraphics[scale=0.56]{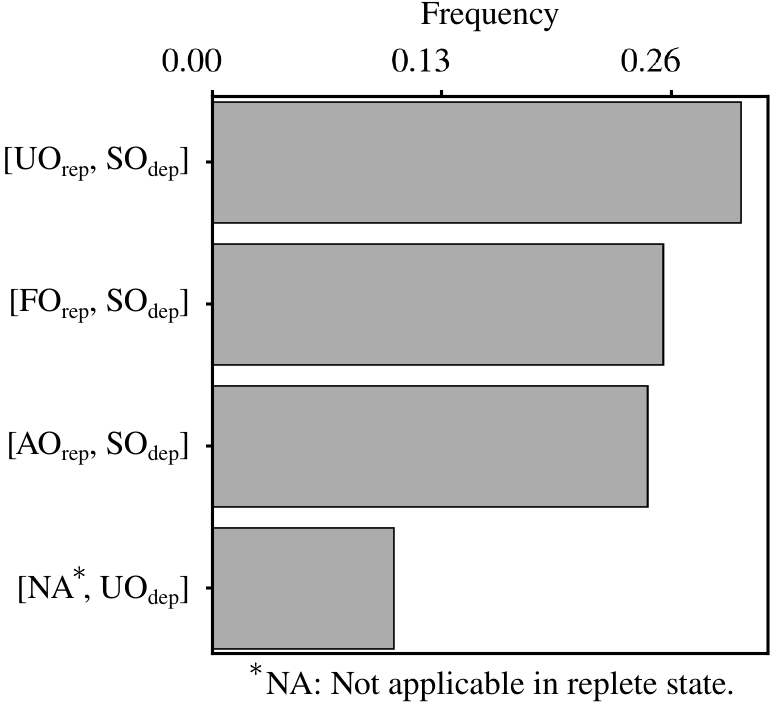}
	\caption{Histogram illustrating the distribution of outcome profiles, defined by the pair of outcomes in the replete and depleted resource states, for the transition vector ${\bm \tau}_{0010}$. The frequency of each outcome profile is obtained by summing the stationary probabilities of population states that produce the same pair of outcomes in the replete and the depleted states. Only profiles with frequencies exceeding $20\%$ of the most abundant profile are shown. Parameters are fixed at $N_o=N_a=100$, $\delta=0.99$, and $w=0.5$ for illustrative purpose.}
	\label{fig:bar_plot_tau0100}
\end{figure}

Using the aforementioned convention, if a strategy pair exclusively generates the action pair (H, H) in the replete state and the action pair (L, H) in the depleted state, then this scenario is conveniently denoted by $[\mathrm{FO}_{\mathrm{rep}},\mathrm{SO}_{\mathrm{dep}}]$---an \emph{outcome profile}. We then count the number of strategy pairs in each outcome profile and determine the corresponding frequency by summing the asymptotic frequencies of all strategy pairs belonging to that outcome profile. Finally, we present them in a histogram and display only those outcome profiles whose frequencies exceed 20$\%$ of the frequency of the most abundant outcome profile; see Fig.~\ref{fig:bar_plot_tau0100}.

 A common pattern  emerges from Fig.~\ref{fig:bar_plot_tau0100}: spiteful outcome in the depleted state is highly prevalent. This is because spiteful outcome prevents further exploitation of the resource and allows it to transition to the replete state, thereby creating opportunities for larger future payoffs. In particular, the outcome profile characterized by a fair outcome in the replete state and a spiteful outcome in the depleted state attains quartile frequency. Fair outcome in the replete state occurs due to repeated interactions. To exploit the abundant resource, the offerer makes a high offer to avoid unsuccessful agreements, while the accepter demands a high share to build a reputation as a tough accepter by rejecting unfair offers in the replete state.  Consequently, fair outcomes emerge in the replete state. Evidently, fair outcomes in turn deplete the resource. The depleted resource then favors spiteful outcomes which facilitate resource replenishment. Once replenished, the resource again promotes fair behavior. Thus, a self-sustaining spite driven resource feedback loop emerges in the stochastic mini-UG. This loop does not arise for the transition vectors ${\bm \tau}_{1111}$, since repeated interactions occur predominantly in the replete states. Moreover, spiteful outcomes do not play a distinct role in the resource transition for this transition vector.

\section{Discussion}
We have introduced a stochastic game framework within the paradigm of the UG, incorporating ownership---an omnipresent phenomenon in biological systems~\cite{HRB_2016_JEB}---to investigate the relationship between the resource and spiteful and fair behaviors. To make the framework the simplest non-trivial one, we considered a two-state resource in which repeated interactions are governed by the mini-UG. In this framework, owner acts as offerer who repeatedly interacts with an accepter to negotiate an agreement for resource exploitation, and the accepter demands a share in return for her expertise in resource exploitation. A successful agreement enables resource exploitation, whereas spiteful behavior hinders agreement, thereby limiting exploitation and allowing the resource to replenish.

Since the underlying game of repeated interaction is assumed to be the mini-UG, the offerer proposes, and the accepter demands, either an equal share or a share close to zero. Furthermore, we assume that the offerer and the accepter employ reactive strategies: the offerer's offer is contingent on the accepter’s most recent demand, and the accepter's demand is contingent on the offerer’s most recent offer. We then examined the evolution of spite and fairness under a mutation--selection process in a two-species finite well-mixed population consisting of offerer and accepter subpopulations.

The emergence of spiteful and fair behaviours largely depends on the growth rate of the resource. Fairness is sustained, whereas the level of spite remains low for resources with high growth rates. However, at low growth rates, an interesting feedback mechanism involving the resource, spite, and fairness emerges. Spiteful behaviour dominates in the depleted state, which helps the resource recover to the replete state, since it hinders agreement. Then, the repeated interactions between them foster fairness in the replete state, and fairness in turn depletes the resource as fair behavior facilitates agreement. In essence, a spite and fairness driven resource feedback loop persists in the stochastic mini-UG for resources with low growth rates.

Experimental studies have shown that resource scarcity can trigger spiteful behaviour~\cite{PVH_2014_JPE}, whereas repeated interactions promote fairness~\cite{S_1999_Experimental}. In addition, ownership has been identified as an important factor in resisting resource depletion~\cite{CO_2010_EJ}. Incorporating these key factors, our model theoretically yields qualitatively analogous outcomes: spiteful behavior predominantly emerges in the depleted resource state, while fairness prevails in the abundant replete state due to repeated interactions. Furthermore, existing game--environment feedback frameworks are largely based on the paradigm of cooperation~\cite{HSC_2018_NAT}. Our framework extends this direction to the UG setting and may facilitate future studies on the role of resource competition and environmental feedback in shaping the evolution of social and anti-social behaviors.

While organisms compete for resources, information about the resource state can substantially influence their strategic interactions. In reality, acquiring complete information is often difficult and impractical, and organisms may also make errors in perception due to limitations in their sensory apparatus. In contrast, our model assumes that individuals possess complete information about the resource state. Exploring interactions under incomplete information may reveal the role of information in shaping the evolution of spite, fairness, and resource dynamics~\cite{KHS_2023_NATCOM, PBC_2025_PRE}. Moreover, cognitively advanced organisms may store longer interaction histories. In particular, \emph{Homo economicus}~\cite{Doucouliagos1994} individuals modulate behaviour through inference and learning. Learning is also found to promote fair behaviour in experimental settings~~\cite{Roth1995}. Introducing more sophisticated Markovian strategies and learning mechanisms into our framework therefore poses a potential direction for future research~\cite{PD_2012_PNAS, PNH_2022_NATCOM, LHN_2023_PLoS, IT_1965_EBV, CH_1999_E, MF_2002_PNAS, PSC_2026_PRE}.

\section*{Data availability statement}
All the relevant numerical codes used to generate the results of this paper are publicly available on GitHub:~\href{https://github.com/ArunavaHub/StochasticUG.git}{\url{https://github.com/ArunavaHub/StochasticUG.git}}.
\acknowledgements
University Grant Commission (India) is acknowledged for awarding a Senior Research Fellowship to PM.  AP acknowledges the support from the Indian Institute of Technology Kanpur in the form of FARE Fellowship.

\appendix

\section{Fairness and spite rate of reactive strategies in two-player stochastic mini-UG}
\label{appen:Fairness and spite rate of reactive strategy in two-player stochastic mini-UG}
We present an analytical derivation of the spite and fairness rates, $\nu(\bm{S}_o, \bm{S}_a)$ and $\gamma(\bm{S}_o, \bm{S}_a)$, for reactive strategies in a two-player stochastic mini-UG using a Markov chain approach. In our case, each Markov chain state corresponds to a resource state and an action pair of the offerer and the accepter. These states are represented by $\omega = (s^i, \text{a}_o, \text{a}_a)$, where $s^i \in \{s^1, s^2\}$ and $\text{a}_o, \text{a}_a \in \{\text{H}, \text{L}\}$. The transition probabilities between states are determined by the transition vector and the reactive strategies of the offerer and the accepter. Let $m_{\omega, \omega'}$ denote the probability of transitioning from the current state $\omega = (s^i, \text{a}_o, \text{a}_a)$ to the next state $\omega' = (s^{i'}, \text{a}'_o, \text{a}'_a)$. Then, it is given by
\begin{equation}
	m_{\omega, \omega'} = z_r \cdot z_o \cdot z_a.
\end{equation}

The component $z_r$ represents the probability of transitioning the resource state from $s^i$ to $s^{i'}$. It is determined by the transition vector $\tau$ as follows
\begin{equation}
	z_r =
	\begin{cases}
		\tau^i_{\text{a}_o \text{a}_a}, & \text{if } s^{i'} = s^1, \\
		1 - \tau^i_{\text{a}_o \text{a}_a}, & \text{if } s^{i'} = s^2.
	\end{cases}
\end{equation}
The remaining components, $z_o$ and $z_a$, are determined by the reactive strategies of the offerer and the accepter, respectively. For the offerer,
\begin{equation}
	z_o =
	\begin{cases}
		p^i_o, & \text{if } \text{a}'_o = \text{H} \text{ and } \text{a}_a = \text{H}, \\
		1 - p^i_o, & \text{if } \text{a}'_o = \text{L} \text{ and } \text{a}_a = \text{H}, \\
		q^i_o, & \text{if } \text{a}'_o = \text{H} \text{ and } \text{a}_a = \text{L}, \\
		1 - q^i_o, & \text{if } \text{a}'_o = \text{L} \text{ and } \text{a}_a = \text{L}.
	\end{cases}
\end{equation}
Likewise, for the accepter,
\begin{equation}
	z_a =
	\begin{cases}
		p^i_a, & \text{if } \text{a}'_a = \text{H} \text{ and } \text{a}_o = \text{H}, \\
		1 - p^i_a, & \text{if } \text{a}'_a = \text{L} \text{ and } \text{a}_o = \text{H}, \\
		q^i_a, & \text{if } \text{a}'_a = \text{H} \text{ and } \text{a}_o = \text{L}, \\
		1 - q^i_a, & \text{if } \text{a}'_a = \text{L} \text{ and } \text{a}_o = \text{L}.
	\end{cases}
\end{equation}
The above procedure yields all transition probabilities $m_{\omega, \omega'}$, which together define the Markov chain transition matrix ${\sf M}(\bm{S}_o, \bm{S}_a)$.

We next derive the average state vector from the transition matrix. To do so, an initial state vector must be specified. We denote the initial state vector by $\bm{\sigma}^{(0)}$ and each element ${\sigma}^{(0)}_{s^i \text{a}_o \text{a}_a}$ of $\bm{\sigma}^{(0)}$ represents the probability of finding the system initially in the state $(s^i, \text{a}_o, \text{a}_a)$. Since the initial resource state is taken to be depleted, we assign the probability one to a particular component ${\sigma}^{(0)}_{s^2 \text{a}_o \text{a}_a}$ corresponding to the depleted state, depending on the pure strategies of the offerer and the accepter. Given the initial state vector $\bm{\sigma}^{(0)}$, the weighted average state vector $\bar{\bm{\sigma}}$ is given by
\begin{equation}
	\bar{\bm{\sigma}}=\frac{\sum_{k=1}^{\infty} \bm{\sigma}^{(0)}(\delta{\sf M})^{k-1}}{\sum_{k=1}^{\infty}\delta^{k-1}}=(1-\delta)\bm{\sigma}^{(0)}({\sf I}-\delta{\sf M})^{-1}.
	\label{eq:wighted_averaged_state_vector}
\end{equation}
where ${\sf I}$ is the $8 \times 8$ identity matrix. Each element $\bar{\sigma}_{s^i \text{a}_o \text{a}_a}$ of $\bar{\bm{\sigma}}$ represents the probability of finding the system in the state $(s^i, \text{a}_o, \text{a}_a)$ over the effective length of the game. 

The spite and fairness rates can now be obtained directly from the weighted average state vector as follows:
\begin{eqnarray}
	\nu(\bm{S}_o, \bm{S}_a) &=& \bar{\sigma}_{s^1\text{LH}} + \bar{\sigma}_{s^2\text{LH}}, 
	\label{eq:spite_rate}\\
	\gamma(\bm{S}_o, \bm{S}_a) &=& \bar{\sigma}_{s^1\text{HH}} + \bar{\sigma}_{s^2\text{HH}}.
	\label{eq:fairness_rate}
\end{eqnarray}
The probability of the resource being in the replete state can also be obtained from the weighted state vector using the following relation:
\begin{eqnarray}
	\beta(\bm{S}_o, \bm{S}_a) &=& \bar{\sigma}_{s^1\text{HH}} + \bar{\sigma}_{s^1\text{HL}}+\bar{\sigma}_{s^1\text{LH}} + \bar{\sigma}_{s^1\text{LL}}. 
	\label{eq:spite_rate}
\end{eqnarray}
The components of the average state vector depend on $(\bm{S}_o, \bm{S}_a)$; for notational simplicity, we write $\sigma_{s^i \text{a}_o \text{a}_a}$ in place of $\sigma_{s^i \text{a}_o \text{a}_a}(\bm{S}_o, \bm{S}_a)$. The expected payoffs of the offerer and the accepter, as defined in Eqs.~\ref{eq:Average_cum_payoff_offerer} and~\ref{eq:Average_cum_payoff_accepter}, can be expressed in terms of the weighted average state vector $\bar{\bm{\sigma}}$ as follows:
\begin{eqnarray} 
	\pi_o(\bm{S}_o, \bm{S}_a) &=& (1-h)\left(\bar{\sigma}_{s^1\text{HH}} + \bar{\sigma}_{s^1\text{HL}}\right) + (1-l)\bar{\sigma}_{s^1\text{LL}} \nonumber\\
	&& + n(1-h)\left(\bar{\sigma}_{s^2\text{HH}} + \bar{\sigma}_{s^2\text{HL}}\right) + n(1-l)\bar{\sigma}_{s^2\text{LL}}, \nonumber \\
	\pi_a(\bm{S}_o, \bm{S}_a) &=& h\left(\bar{\sigma}_{s^1\text{HH}} + \bar{\sigma}_{s^1\text{HL}}\right) + l\bar{\sigma}_{s^1\text{LL}} \nonumber\\
	&& + nh\left(\bar{\sigma}_{s^2\text{HH}} + \bar{\sigma}_{s^2\text{HL}}\right) + nl\bar{\sigma}_{s^2\text{LL}}.
	\label{eq:wighted_sum_payoff} 
\end{eqnarray}

\section{Analytical framework of two-species long-term stochastic evolutionary dynamics}
\label{appen:Analytical framework of two-species long-term stochastic evolutionary dynamics}
We illustrate the analytical framework of long-term mutation-selection process in two-species population using Markov chain method to find $\alpha^{(t)}_{(\bm{S}_o, \bm{S}_a)}$  required to calculate the average fairness and spite rate in the population over time. Therefore, we consider a well mixed population consists of the offerer and the accepter subpopulations with population sizes $N_o$ and $N_a$, and the population is subjected to rare mutation process.

As the population is consist of two subpopulations, the mutation can occur in the system various ways in each time step. Either any one subpopulation undergoes mutation and it either fixates or goes extinct, or both subpopulations together experience mutation. When both subpopulations face the rare mutant, there are three possible outcomes in the next time step: the mutant in one of two subpopulation fixates or the mutations in both subpopulations fixate, or the mutation goes extinct from both subpopulation. Note that both subpopulations are monomorphic before and after fixation or extinction of rare mutant. A new population state emerges only upon the fixation of the rare mutant.


Thus, the evolution of the population state is completely governed by the rare mutation process. As the state of the population is characterized by the pair of strategies of offerer and accepter, and the possible number pure strategies of the offerer and the accepter are 16, the number of possible population states is 256. The transition between these population states completely described by discrete-time discrete-states Markov chain. Clearly, the population states are the states of the Markov chain where the transition probability from one state to another state is determined by the fixation probability and the rate of mutation process.

Unlike the mutation-selection process in homogeneous population, the mutant of various types in a same resident population can end up in the same future population state. For example, the transition from a current population $((\text{F}_o; \text{C}_o), (\text{F}_a; \text{C}_a))$ to future population state $((\text{U}_o; \text{U}_o),(\text{F}_a; \text{C}_a))$ can be governed by 16 mutant pairs which are $((\text{U}_o; \text{U}_o), \bm{S}_a)~\forall \bm{S}_a\in\mathbb{S}_a$. In this case, the mutant in the offerer population needs to fixate whereas the mutant in the accepter population must go for extinction. Therefore, the transition probability for the transition like from state $(\bm{S}_o, \bm{S}_a)$ to $(\bm{S}_o, \tilde{\bm{S}}_a)$ or from the state $(\bm{S}_o, \bm{S}_a)$ to $(\tilde{\bm{S}}_o, \bm{S}_a)$ is determined by the fixation probabilities of all mutant pairs that lead the transition. The detailed discussion of fixation probability in two-species population is provided in Appendix~\ref{appen:Birth--death process in two-species population}. 

However, the transition in which the strategies of the both subpopulation are changed in the future population state can be achieved when mutants occur in both subpopulations. For example, the transition from the population state $((\text{F}_o; \text{C}_o), (\text{F}_a; \text{C}_a) )$ to $((\text{U}_o; \text{U}_o),(\text{U}_a; \text{U}_a))$ can occur for only one mutant pair $((\text{U}_o; \text{U}_o),(\text{U}_a; \text{U}_a))$. Therefore, the transition probability for the transition like from the state $(\bm{S}_o, \bm{S}_a)$ to $(\tilde{\bm{S}}_o, \tilde{\bm{S}}_a)$ is determined by the fixation probability of the mutant pair $(\tilde{\bm{S}}_o, \tilde{\bm{S}}_a)$.

To formalize the transition probabilities, let $\rho_{\bm{S}'_o \bm{S}'_a}((\bm{S}_o, \bm{S}_a), (\tilde{\bm{S}}_o, \tilde{\bm{S}}_a))$ denote the fixation probability of a mutant pair $(\tilde{\bm{S}}_o, \tilde{\bm{S}}_a)$ arising in a resident population with state $(\bm{S}_o, \bm{S}_a)$, resulting in a subsequent population state $(\bm{S}'_o, \bm{S}'_a)$, where $\bm{S}'_o \in \{\bm{S}_o, \tilde{\bm{S}}_o\}$ and $\bm{S}'_a \in \{\bm{S}_a, \tilde{\bm{S}}_a\}$. For example, $\rho_{\bm{S}_o \tilde{\bm{S}}_a}((\bm{S}_o, \bm{S}_a), (\tilde{\bm{S}}_o, \tilde{\bm{S}}_a))$ corresponds to the probability that the mutant pair $(\tilde{\bm{S}}_o, \tilde{\bm{S}}_a)$ drives the population from $(\bm{S}_o, \bm{S}_a)$ to $(\bm{S}_o, \tilde{\bm{S}}_a)$. Furthermore, let $\mu_o$ and $\mu_a$ denote the probabilities of mutation in the offerer and the accepter subpopulations, respectively, at each time step.

We now turn to the construction of the Markov chain transition matrix by specifying the transition probabilities. Note that the population can transition from $(\bm{S}_o, \bm{S}_a)$ to one of three states: $(\tilde{\bm{S}}_o, \bm{S}_a)$, $(\bm{S}_o, \tilde{\bm{S}}_a)$, or $(\tilde{\bm{S}}_o, \tilde{\bm{S}}_a)$. We below analyze these cases separately.

\begin{itemize}
	
	\item The transition probability from $(\bm{S}_o, \bm{S}_a)$ to $(\tilde{\bm{S}}_o, \bm{S}_a)$ is given by
	\begin{eqnarray*}
		r[(\bm{S}_o, \bm{S}_a), (\tilde{\bm{S}}_o, \bm{S}_a)] &=& \mu_o(1-\mu_a)\rho_{\tilde{\bm{S}}_o \bm{S}_a}[(\bm{S}_o, \bm{S}_a), (\tilde{\bm{S}}_o, \bm{S}_a)] \\
		&+& \mu_o\mu_a \sum_{\substack{\bm{S}'_a \neq \bm{S}_a \\ \bm{S}'_a \in \mathbb{S}_a}} 
		\rho_{\tilde{\bm{S}}_o \bm{S}_a}[(\bm{S}_o, \bm{S}_a), (\tilde{\bm{S}}_o, \bm{S}'_a)].
	\end{eqnarray*}
	The first term captures the scenario in which a mutation arises only in the offerer subpopulation, while no mutation occurs in the accepter subpopulation. The second term accounts for the case where mutations arise in both subpopulations, but fixation occurs only in the offerer subpopulation. It may be noted that $\mu_o\mu_a < \mu_o(1-\mu_a)$ since $\mu_o, \mu_a \ll 1$ highlighting the fact that the simultaneous mutations is rarer than single mutation in only one subpopulation.
	
	\item The transition probability from $(\bm{S}_o, \bm{S}_a)$ to $(\bm{S}_o, \tilde{\bm{S}}_a)$ is analogously given by
	\begin{eqnarray*}
		r[(\bm{S}_o, \bm{S}_a), (\bm{S}_o, \tilde{\bm{S}}_a)] &=& (1-\mu_o)\mu_a \rho_{\bm{S}_o \tilde{\bm{S}}_a}[(\bm{S}_o, \bm{S}_a), (\bm{S}_o, \tilde{\bm{S}}_a)] \\
		&+& \mu_o\mu_a \sum_{\substack{\bm{S}'_o \neq \bm{S}_o \\ \bm{S}'_o \in \mathbb{S}_o}} 
		\rho_{\bm{S}_o \tilde{\bm{S}}_a}[(\bm{S}_o, \bm{S}_a), (\bm{S}'_o, \tilde{\bm{S}}_a)].
	\end{eqnarray*}
	
	\item A transition from $(\bm{S}_o, \bm{S}_a)$ to $(\tilde{\bm{S}}_o, \tilde{\bm{S}}_a)$ occurs only when mutations arise and fixate in both subpopulations. The corresponding transition probability is therefore
	\begin{eqnarray*}
		r[(\bm{S}_o, \bm{S}_a), (\tilde{\bm{S}}_o, \tilde{\bm{S}}_a)] &=& \mu_o\mu_a \, \rho_{\tilde{\bm{S}}_o \tilde{\bm{S}}_a}[(\bm{S}_o, \bm{S}_a), (\tilde{\bm{S}}_o, \tilde{\bm{S}}_a)].
	\end{eqnarray*}
	
	\item The probability of remaining in the same state $(\bm{S}_o, \bm{S}_a)$ is therefore
	\begin{eqnarray*}
		r[(\bm{S}_o, \bm{S}_a), (\bm{S}_o, \bm{S}_a)] &=& 1 - r[(\bm{S}_o, \bm{S}_a), (\tilde{\bm{S}}_o, \bm{S}_a)] \\
		&& - r[(\bm{S}_o, \bm{S}_a), (\bm{S}_o, \tilde{\bm{S}}_a)] \\
		&& - r[(\bm{S}_o, \bm{S}_a), (\tilde{\bm{S}}_o, \tilde{\bm{S}}_a)].
	\end{eqnarray*}
	
\end{itemize}
One can construct a $256$-dimensional Markov transition matrix, denoted by $\sf R$, from the above transition probabilities. Suppose $\bm{\alpha}(t)$ be the probability row vector at time $t$, then $\bm{\alpha}(t)$ given an initial distribution $\bm{\alpha}(0)$ is found from the Markov chain relation
$
\bm{\alpha}(t) = \bm{\alpha}(0)\, {\sf R}^t.
$
The component $\alpha^{(t)}_{(\bm{S}_o, \bm{S}_a)}$ of $\bm{\alpha}(t)$ gives the abundance of the population state $(\bm{S}_o, \bm{S}_a)$ at time $t$. In our case, we assume the initial population state is $((\text{U}_o; \text{U}_o),(\text{U}_a; \text{U}_a))$. Accordingly, the component of $\bm{\alpha}(0)$ corresponding to $((\text{U}_o; \text{U}_o),(\text{U}_a; \text{U}_a))$ is set to $1$, while all other components are set to $0$.
\section{Birth--death process in two-species population}
\label{appen:Birth--death process in two-species population}
We now determine fixation probabilities from a birth--death process in a heterogeneous population consisting of the offerer and the accepter subpopulations of sizes $N_o$ and $N_a$, respectively. Each subpopulation has two possible strategies: $\{\bm{S}_o, \tilde{\bm{S}}_o\}$ for the offerer and $\{\bm{S}_a, \tilde{\bm{S}_a}\}$ for the accepter. Let $n_o$ and $n_a$ denote the numbers of individuals adopting strategies $\tilde{\bm{S}}_o$ and $\tilde{\bm{S}}_a$ in the offerer and the accepter subpopulations, respectively. The population state is then described by the pair $(n_o, n_a)$. 

Since an offerer—regardless of whether it adopts $\bm{S}_o$ or $\tilde{\bm{S}}_o$—interacts with $n_a$ accepters using $\tilde{\bm{S}_a}$ and $N_a - n_a$ accepters using  $\bm{S}_a$, the expected payoffs of strategies $\bm{S}_o$ and $\tilde{\bm{S}}_o$ can be found as follows
\begin{eqnarray*}
	E_{\bm{S}_o}(n_a) &=& \frac{1}{N_a}\left[n_a\,\pi_o(\bm{S}_o, \bm{S}_a) + (N_a - n_a)\,\pi_o(\bm{S}_o, \tilde{\bm{S}}_a)\right],\\
	E_{\tilde{\bm{S}}_o}(n_a) &=& \frac{1}{N_a}\left[n_a\,\pi_o(\tilde{\bm{S}}_o, \bm{S}_a) + (N_a - n_a)\,\pi_o(\tilde{\bm{S}}_o, \tilde{\bm{S}}_a)\right].
\end{eqnarray*}
Analogously, an accepter adopting strategy $\bm{S}_a$ or $\tilde{\bm{S}}_a$ interacts with $n_o$ offerers using $\tilde{\bm{S}}_o$ and $N_o - n_o$ offerers using $\bm{S}_o$. The corresponding expected payoffs of strategies $\bm{S}_a$ and $\tilde{\bm{S}}_a$ are therefore given by
\begin{eqnarray*}
	E_{\bm{S}_a}(n_o) &=& \frac{1}{N_o}\left[n_o\,\pi_a(\bm{S}_o, \bm{S}_a) + (N_o - n_o)\,\pi_a(\tilde{\bm{S}}_o, \bm{S}_a)\right],\\
	E_{\tilde{\bm{S}}_a}(n_o) &=& \frac{1}{N_o}\left[n_o\,\pi_a(\bm{S}_o, \tilde{\bm{S}}_a) + (N_o - n_o)\,\pi_a(\tilde{\bm{S}}_o, \tilde{\bm{S}}_a)\right].
\end{eqnarray*}

It is usually considered in the literature that the fitness of a player is to be an exponential function of the expected payoff. Accordingly, the fitness of an offerer adopting strategy $\bm{S}_o$ or $\tilde{\bm{S}}_o$ is given by
\[
f_{\bm{S}_o} = e^{w E_{\bm{S}_o}}, \qquad f_{\tilde{\bm{S}}_o} = e^{w E_{\tilde{\bm{S}}_o}},
\]
and, similarly, the fitness of an accepter adopting strategy $\bm{S}_a$ or $\tilde{\bm{S}}_a$ is
\[
f_{\bm{S}_a} = e^{w E_{\bm{S}_a}}, \qquad f_{\tilde{\bm{S}}_a} = e^{w E_{\tilde{\bm{S}}_a}}.
\]
Here, $w$ denotes the strength of selection, which determines the extent to which payoffs contribute to fitness. In principle, it can take values from zero to infinity. When its value approaches to zero, random drift dominates in the fitness.

These fitness drives the birth--death process in the two-species population. In this process, at each time step, an individual from each subpopulation is chosen to reproduce with a probability proportional to fitness of the individual. Independently, another individual from each subpopulation is chosen uniformly at random to die. The offspring then replaces the death individual within the same subpopulation, thereby keeping the population sizes constant.

An offerer with strategy $\bm{S}_o$ or $\tilde{\bm{S}}_o$ is drawn for reproduction with probabilities
\[
p^r_{\bm{S}_o} = \frac{(N_o - n_o) f_{\bm{S}_o}}{\bar{f}_o}, 
\qquad
p^r_{\tilde{\bm{S}}_o} = \frac{n_o f_{\tilde{\bm{S}}_o}}{\bar{f}_o}
\]
respectively, where $\bar{f}_o = (N_o - n_o) f_{\bm{S}_o} + n_o f_{\tilde{\bm{S}}_o}$ denotes the total fitness of the offerer subpopulation. Analogously, for the accepter subpopulation, the corresponding probabilities for the strategies $\bm{S}_a$ and $\tilde{\bm{S}}_a$ are
\[
p^r_{\bm{S}_a} = \frac{(N_a - n_a) f_{\bm{S}_a}}{\bar{f}_a}, 
\qquad
p^r_{\tilde{\bm{S}}_a} = \frac{n_a f_{\tilde{\bm{S}}_a}}{\bar{f}_a},
\]
where $\bar{f}_a = (N_a - n_a) f_{\bm{S}_a} + n_a f_{\tilde{\bm{S}}_a}$ is the total fitness of the accepter subpopulation. Here, the superscript $r$ represents the probability that a player is chosen for reproduction.

Since individuals are chosen uniformly at random for death, the probability that a player of a given type is selected for death equals its frequency in the population. Thus, the probabilities that an offerer with strategy $\tilde{\bm{S}}_o$ and an accepter with strategy $\tilde{\bm{S}}_a$ are chosen for death are
\[
p^d_{\tilde{\bm{S}}_o} = \frac{n_o}{N_o}, 
\qquad
p^d_{\tilde{\bm{S}}_a} = \frac{n_a}{N_a}.
\]
Accordingly, an offerer using strategy $\bm{S}_o$ and an accepter using strategy $\bm{S}_a$ are chosen to die with probabilities $p^d_{\bm{S}_o} = 1 - p^d_{\tilde{\bm{S}}_o}$ and $p^d_{\bm{S}_a} = 1 - p^d_{\tilde{\bm{S}}_a}$, respectively. Here, the superscript $d$ denotes the probability of being selected for death.

We simulate the birth--death process using these probabilities to estimate fixation probabilities numerically. Depending on the quantity of interest, mutations may be introduced in one or both subpopulations. Consider a mutant pair $(\tilde{\bm{S}}_o, \tilde{\bm{S}}_a)$ appearing in a resident population $(\bm{S}_o, \bm{S}_a)$. In this case, the initial population state is $(n_o, n_a) = (1, 1)$. From this initial state, the process can be absorbed into one of three absorbing states: $(N_o, 0)$, $(0, N_a)$, or $(N_o, N_a)$. We therefore simulate the stochastic process from the initial state $(1,1)$ over many independent realizations. The fixation probabilities $\rho_{\tilde{\bm{S}}_o \bm{S}_a}[(\bm{S}_o, \bm{S}_a), (\tilde{\bm{S}}_o, \tilde{\bm{S}}_a)]$, $\rho_{\bm{S}_o \tilde{\bm{S}}_a}[(\bm{S}_o, \bm{S}_a), (\tilde{\bm{S}}_o, \tilde{\bm{S}}_a)]$, and $\rho_{\tilde{\bm{S}}_o \tilde{\bm{S}}_a}[(\bm{S}_o, \bm{S}_a), (\tilde{\bm{S}}_o, \tilde{\bm{S}}_a)]$ are then estimated from the frequencies with which the process is absorbed into $(N_o, 0)$, $(0, N_a)$, and $(N_o, N_a)$ respectively.

When a mutation arises in only one subpopulation, the relevant fixation probabilities are $\rho_{\tilde{\bm{S}}_o \bm{S}_a}[(\bm{S}_o, \bm{S}_a), (\tilde{\bm{S}}_o, \bm{S}_a)]$ and $\rho_{\bm{S}_o \tilde{\bm{S}}_a}[(\bm{S}_o, \bm{S}_a), (\bm{S}_o, \tilde{\bm{S}}_a)]$. For a mutant of type $(\tilde{\bm{S}}_o, \bm{S}_a)$, the fixation probability $\rho_{\tilde{\bm{S}}_o \bm{S}_a}[(\bm{S}_o, \bm{S}_a), (\tilde{\bm{S}}_o, \bm{S}_a)]$ is given by the fraction of realizations that are absorbed in the state $(N_o, 0)$, starting from the initial condition $(n_o, n_a) = (1, 0)$. Similarly, for a mutant of type $(\bm{S}_o, \tilde{\bm{S}}_a)$, the fixation probability $\rho_{\bm{S}_o \tilde{\bm{S}}_a}[(\bm{S}_o, \bm{S}_a), (\bm{S}_o, \tilde{\bm{S}}_a)]$ corresponds to the fraction of realizations absorbed in $(0, N_a)$, starting from $(n_o, n_a) = (0, 1)$.

In both cases, the birth--death process effectively operate within a single subpopulation, as mutations arise in only one subpopulation. Consequently, these scenarios reduce to standard birth--death processes in homogeneous populations, which allow us to use known analytical expressions of fixation probabilities in homogeneous population with the necessary adjustments.

\bibliography{Patra_manuscript_et_al}
\end{document}